\documentclass{WileyMSP-template}
 \usepackage[numbers,sort&compress]{natbib}
\usepackage{siunitx}
\usepackage[table]{xcolor}
\usepackage{makecell}
\usepackage{amsmath}
\usepackage{placeins}
\usepackage{caption}
\usepackage{ragged2e}

\begin{document}

\pagestyle{fancy}
\rhead{}
\title{Bandwidth-Tunable Quantum Light Source at \SI{1.5} {\micro\meter}}

\maketitle
         
\author{Wei Wu}$^{1,2,3,4}$
\author{Yun-Ru Fan}$^{2,3,4,\dagger}$
\author{Ri-Yao Song}$^{2,3,4}$
\author{Si Shen}$^{5}$
\author{Zi-Chang Zhang}$^{5}$
\author{Hai-Zhi Song}$^{2,5}$
\author{Hao Li}$^{6}$
\author{Li-Xing You}$^{6}$
\author{Ping-He Wang}$^{1,\ddagger}$
\author{Yan-Yu Wei}$^{2}$
\author{Kai Guo}$^{7}$
\author{Guang-Can Guo}$^{2,3,4,8}$
\author{Qiang Zhou}$^{1,2,3,4,8,*}$

\medskip

\begin{affiliations}
$^1$School of Optoelectronic Science and Engineering, University of Electronic Science and Technology of China, Chengdu 611731, China\\
$^2$Institute of Fundamental and Frontier Sciences, University of Electronic Science and Technology of China,
Chengdu 611731, China\\
$^3$Center for Quantum Internet, Tianfu Jiangxi Laboratory, Chengdu 641419, China\\ 
$^4$Key Laboratory of Quantum Physics and Photonic Quantum Information, Ministry of Education, University of Electronic Science and Technology of China, Chengdu 611731, China\\
$^5$Southwest Institute of Technical Physics, Chengdu 610041, China\\
$^6$National Key Laboratory of Materials for Integrated Circuits, Shanghai Institute of Microsystem and Information Technology, Chinese Academy of Sciences, Shanghai 200050, China\\
$^7$Institute of Systems Engineering, AMS, Beijing 100141, China\\
$^8$CAS Center for Excellence in Quantum Information and Quantum Physics, University of Science and Technology of China, Hefei 230026, China\\
$^\dagger$yunrufan@uestc.edu.cn\\
$^\ddagger$wphsci@uestc.edu.cn\\
$^*$zhouqiang@uestc.edu.cn\\
\end{affiliations}

\medskip

\justifying

\begin{abstract}
\noindent
Quantum light sources constitute a crucial physical resource for the construction of quantum networks. Despite remarkable recent progress, there remains a lack of systematic investigation into the bandwidth tunability of quantum light sources under fixed waveguide parameters. In this work, we demonstrate a broadband quantum light source in the \SI{1.5} {\micro\meter} band with tunable bandwidth by changing the temperature of a piece of periodically poled lithium niobate waveguide. In our demonstration, the bandwidth of the quantum light source is tuned from \SI{78.3} {\nano\meter} to \SI{96.2} {\nano\meter} with a temperature change of \SI{1}{\celsius}. Under different bandwidths, the generation rates of correlated photon pairs are greater than \SI{6.3} {\mega\hertz} with coincidence-to-accidental ratios consistently being no less than 608. The energy-time entanglement properties are measured by using the Franson interference with two-photon interference visibilities larger than \num{99.06}\%. Our results provide an effective method for developing the quantum light sources with tunable bandwidth which has great potential for building the large-scale quantum networks.

\end{abstract}

\section{Introduction}

Quantum light sources are key resources in quantum information science, which find widespread applications, including quantum communication \cite{bouwmeester1997experimental,kimble2008quantum,qi202115,shen2023hertz,khodadad2025frequency}, quantum metrology \cite{mitchell2004super,fan2025high,lualdi2025fast}, quantum computing \cite{divincenzo1995quantum,ladd2010quantum,monika2025quantum}. To date, the generation of entangled photon pairs from quantum light sources relies on nonlinear optical processes, including spontaneous four-wave mixing (SFWM) process \cite{reimer2016generation,caspani2017integrated,paesani2020near,chen2025quantum} based on third-order nonlinearity $\chi^{(3)}$, as well as spontaneous parametric down-conversion (SPDC) process \cite{shalm2013three,malik2016multi,zhao2020high,francesconi2023chip,liang2025tunable}, cascaded second-harmonic generation (SHG)/SPDC process \cite{ngah2015ultra,zhang2021high,li2023discrete}, and cascaded sum-frequency generation (SFG)/SPDC process \cite{arahira2020wdm,li2024generation} enabled by second-order nonlinearity $\chi^{(2)}$. Various materials have been used to realize quantum light sources based on the SFWM process, including optical fibers \cite{li2005optical,fang2019three}, silicon \cite{feng2023entanglement,gianini2026silicon}, silicon nitride \cite{lu2019chip,fan2023multi}, gallium nitride \cite{zeng2024quantum}, and silicon carbide \cite{rahmouni2024entangled}, etc. Both the SPDC process in a single waveguide \cite{wu2025broadband} and the SHG/SPDC and SFG/SPDC processes in either two \cite{ngah2015ultra,businger2022non} or a single waveguide \cite{zhang2021high,li2023discrete,li2024generation} have been demonstrated, while the former offers lower noise and higher conversion efficiency. Although various materials, such as potassium titanyl phosphate \cite{evans2010bright,eckstein2011highly}, gallium arsenide \cite{autebert2016integrated,baboux2023nonlinear}, and periodically poled lithium niobate (PPLN) \cite{joshi2018frequency,ma2020ultrabright}, etc., have been developed for realizing the SPDC process, related studies have mainly focused on characterizing the correlation and entanglement properties of quantum light sources and exploring their applications. Under fixed waveguide parameters, systematic investigations into the bandwidth tunability of quantum light sources remain lacking, further limiting the large-scale development of quantum networks \cite{simon2017towards,wengerowsky2018entanglement,wei2022towards,fan2025quantum}.

In this work, we demonstrate a bandwidth-tunable quantum light source at \SI{1.5} {\micro\meter} based on the type-0 SPDC process in a single PPLN waveguide. The source can cover different International Telecommunication Union (ITU) channels on demand. The source exhibits a tunable single photon spectral bandwidth ranging from \SI{78.3} {\nano\meter} to \SI{96.2} {\nano\meter} within a temperature range of \SIrange{43.2}{44.2}{\celsius}, consistent with theoretical calculations. In addition, dense wavelength division multiplexers (DWDMs) are used to select the ITU C55/C59 channels as representative examples to characterize the correlation properties and entanglement quality of the source. The correlated photon-pair generation rates (PGR) remain no less than \SI{6.33} {\mega\hertz}, the coincidence-to-accidental ratios (CAR) remain above \num{608}. The Franson interference visibilities under different phase settings exceed \num{99.06}\%, all of which significantly violate the Bell inequality. This method provides a feasible solution for developing bandwidth-tunable quantum light sources and lays the foundation for the realization of scalable quantum networks.

\section{Principle of Bandwidth Tuning}

This work utilizes correlated photon pairs generated via the type-0 SPDC process in a single PPLN waveguide to realize a bandwidth-tunable quantum light source. The bandwidth of the correlated photon pairs is governed by the phase mismatch $\Delta k$ and the waveguide length. By introducing an appropriate poling period, the phase mismatch $\Delta k$ induced by material dispersion is compensated. Meanwhile, due to the temperature sensitivity of the refractive index, the phase mismatch $\Delta k$ can be tuned by varying the temperature. Therefore, under fixed waveguide parameters, the single photon spectral bandwidth can be tuned through temperature control. In the SPDC process, signal and idler photons can be generated within a certain wavelength range, and the single photon generation rate is given by Equation~\eqref{eq:rate}.

\begin{equation}
R(\omega_{s,i}) \propto \frac{d_{\mathrm{eff}}^2 I_p L^2}{n_p \varepsilon_0 c^3} \cdot \frac{\omega_{i,s} \omega_{s,i}}{n_s^2 n_i^2} \cdot \mathrm{sinc}^2 \left[ \Delta k(\omega_{s,i}, T) \cdot \frac{L}{2} \right]
\label{eq:rate}
\end{equation}

\noindent
Here, $\omega_{s,i}$ are the angular frequencies of the signal or idler photons, $n_p$, $n_s$, $n_i$ are the refractive indices for the pump, signal, and idler, $I_p$ is the pump intensity, $L$ is the interaction length, $\varepsilon_0$ is the vacuum permittivity, $d_{\mathrm{eff}}$ is the effective nonlinear optical coefficient determined by the material properties and geometric parameters, and $T$ represents the temperature. Taking into account the dispersion of the nonlinear medium, the phase mismatch is given by Equation~\eqref{eq:delta_k_temp} \cite{jundt1997temperature}. 

\begin{equation}
\Delta k(\omega_s, \omega_i) = \frac{1}{c_0} \left[ n_p(\omega_p) \omega_p - n_s(\omega_s) \omega_s - n_i(\omega_i) \omega_i \right] - \frac{2\pi}{\Lambda}
\label{eq:delta_k_temp}
\end{equation}

\noindent
Here, $c_0$ is the speed of light in vacuum and $\Lambda$ is the poling period. The refractive index of the nonlinear medium at different wavelengths can be calculated using the Sellmeier equation \cite{jundt1997temperature}, as given by Equation~\eqref{eq:refractive}.

\begin{equation}
n^2 = a_1 + b_1 f + \frac{a_2 + b_2 f}{\lambda^2 - (a_3 + b_3 f)^2} + \frac{a_4 + b_4 f}{\lambda^2 - a_5^2} - a_6 \lambda^2
\quad \text{with} \quad f = (T - 24.5)(T + 570.82)
\label{eq:refractive}
\end{equation}

\begin{figure}[htbp]
  \centering
 \includegraphics[width=1\linewidth]{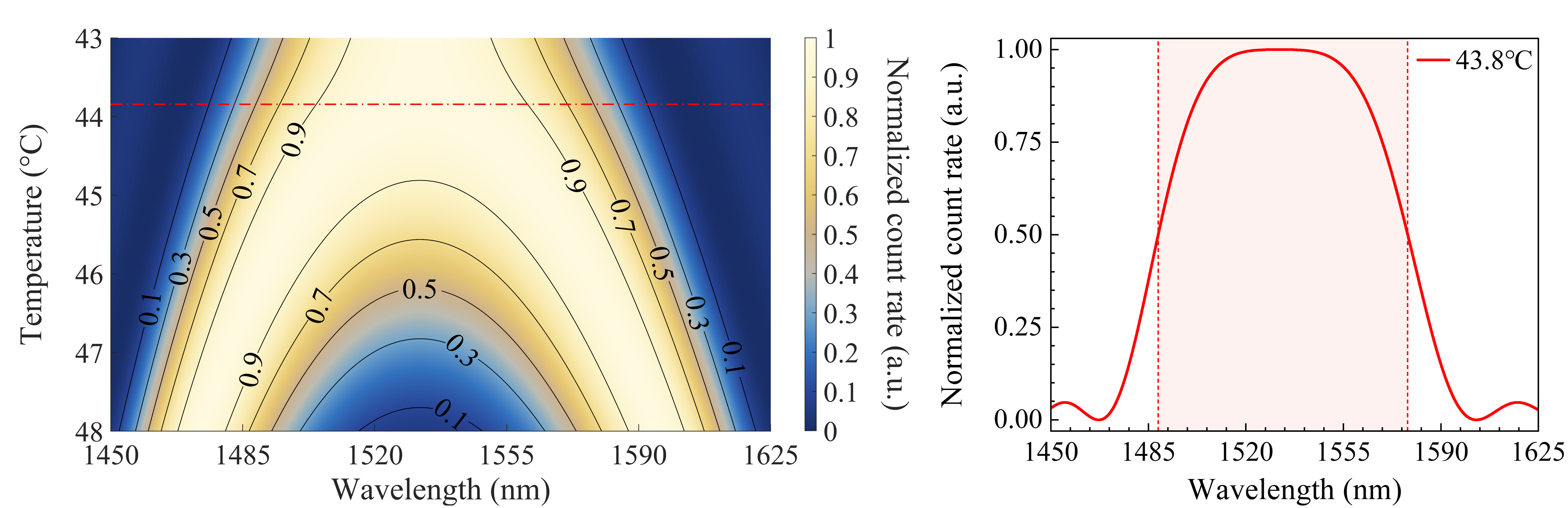}
  \caption{
  Bandwidth tunability of a quantum light source based on a PPLN waveguide.
  a) Temperature-dependent spectral characteristics in the PPLN waveguide. 
  b) Spectral characteristics at \SI{43.8}{\celsius}. 
  }
  \label{fig:fig1}
\end{figure}

\noindent
Here, $\lambda$ denotes the optical wavelength, $f$ is the temperature parameter, and $a_i$ and $b_i$ are the coefficients of the equation taken from reference~\cite{jundt1997temperature}. The coefficients $a_1$ to $a_6$ are \num{5.35583}, \num{0.100473}, \num{0.20692}, \num{100}, \num{11.34927}, and \num{1.5334e-2}, while the coefficients $b_1$ to $b_4$ are \num{4.629e-7}, \num{3.862e-8}, \num{-0.89e-8}, and \num{2.657e-5} \cite{jundt1997temperature}. Based on the above analysis, numerical simulations are performed to map the temperature-dependent single photon spectral properties of correlated photon pairs generated via type-0 SPDC in the PPLN waveguide. The crystal length is set to \SI{20} {\milli\meter}, the pump wavelength is \SI{766.00} {\nano\meter}, and the poling period is \SI{18.4} {\micro\meter}. The single photon wavelength is scanned from \SI{1450} {\nano\meter} to \SI{1625} {\nano\meter}, while the waveguide temperature is varied over the range of \SIrange{43.0}{48.0}{\celsius}. \textbf{Figure~\ref{fig:fig1}} shows the numerical simulation results. The color scale in Figure~\ref{fig:fig1}a represents the normalized count rate of the generated single photons as a function of wavelength and temperature. The red solid curve in Figure~\ref{fig:fig1}b represents the theoretical single photon spectrum at \SI{43.8}{\celsius}.

\FloatBarrier

\section{Characterization of Correlation and Energy-Time Entanglement Properties}

\begin{figure}[h!]
  \centering
  \includegraphics[width=1\linewidth]{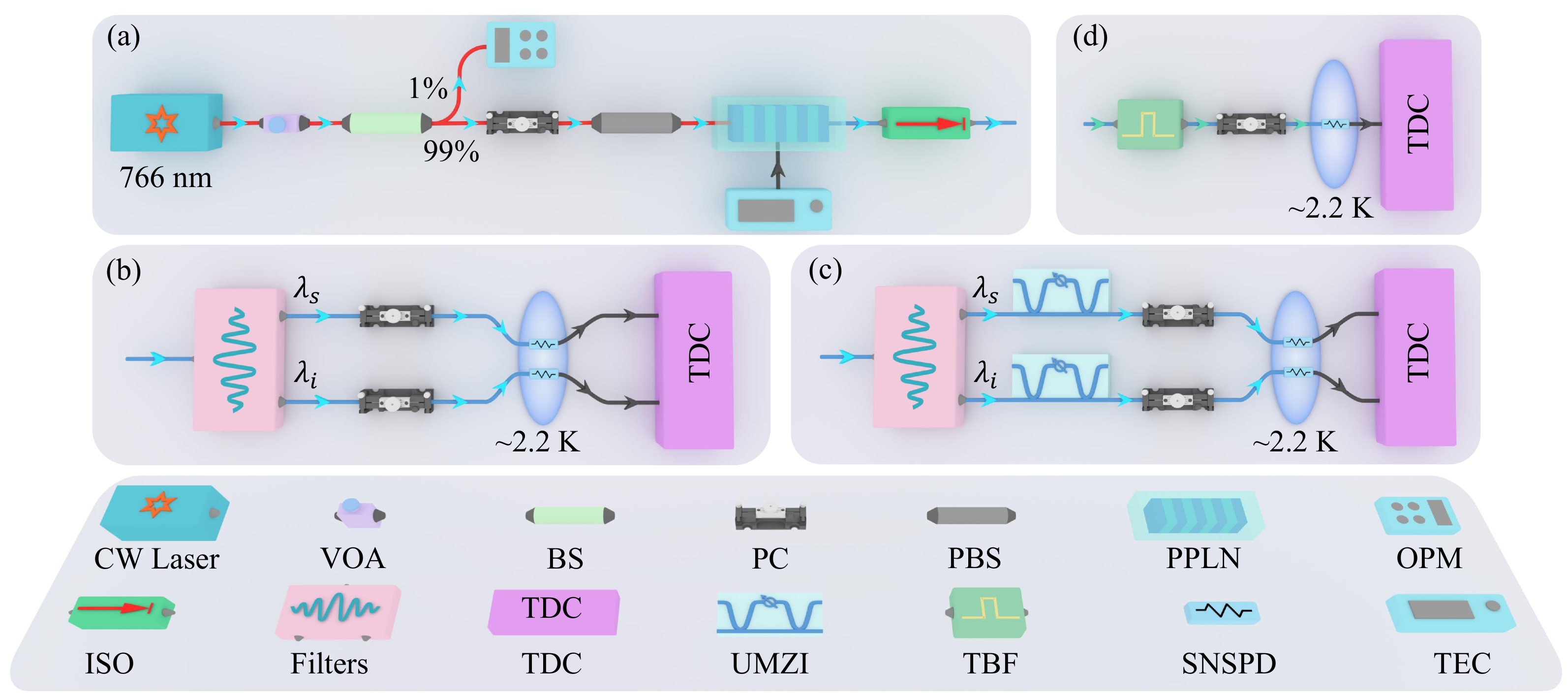} 
  \caption{
    Experimental setup for the preparation and characterization of the quantum light source with tunable bandwidth.
    a) Generation of the correlated photon pairs. 
    b) Correlation properties. 
    c) Energy-time entanglement.
    d) Single photon spectrum. 
    CW Laser, continuous wave laser; VOA, variable optical attenuator; BS, beam splitter; PC, polarization controller; PBS, polarization beam splitter; PPLN, periodically poled lithium niobate; OPM, optical power meter; ISO, optical isolator; Filters, optical filters; TDC, time-to-digital converter; UMZI, unbalanced Mach-Zehnder interferometer; TBF, tunable band-pass filter; SNSPD, superconducting nanowire single photon detector; TEC, thermo-electric controller; $\lambda_s$ and $\lambda_i$ represent the signal and idler photons.
    }
  \label{fig:fig2}
\end{figure}

The experimental setup for the quantum light source with tunable bandwidth is shown in \textbf{Figure~\ref{fig:fig2}}. In our work, a single PPLN waveguide  (Shandong Jiliang)  with a poling period of \SI{18.4} {\micro\meter} and a length of \SI{20} {\milli\meter} is used. The optical setup for correlated photon-pair generation is shown in Figure~\ref{fig:fig2}a. A tunable continuous-wave laser (CW Laser) operating at a center wavelength of \SI{766.00} {\nano\meter} is employed to pump the PPLN waveguide. The pump power is controlled using a variable optical attenuator (VOA) and monitored via a \num{1}\% tap port of a fiber beam splitter (BS) connected to an optical power meter (OPM). The remaining \num{99}\% of the output from the BS is directed to a polarization controller (PC) followed by a polarization beam splitter (PBS), allowing the pump polarization to be adjusted to match the polarization mode required by the PPLN waveguide. Within the PPLN waveguide, correlated photon pairs are generated via the type-0 SPDC process. In this process, a \SI{766.00} {\nano\meter} pump photon is annihilated, spontaneously producing a signal photon and an idler photon at telecom wavelengths. To enable bandwidth tunability of the quantum light source, the waveguide temperature is actively controlled using a benchtop thermo-electric controller (TEC). To eliminate residual pump photons that could contaminate the measurements, an optical isolator (ISO) operated at the telecom band is placed at the output end of the PPLN waveguide.

\FloatBarrier

\begin{figure}[htbp]
  \centering
  \includegraphics[width=1\linewidth]{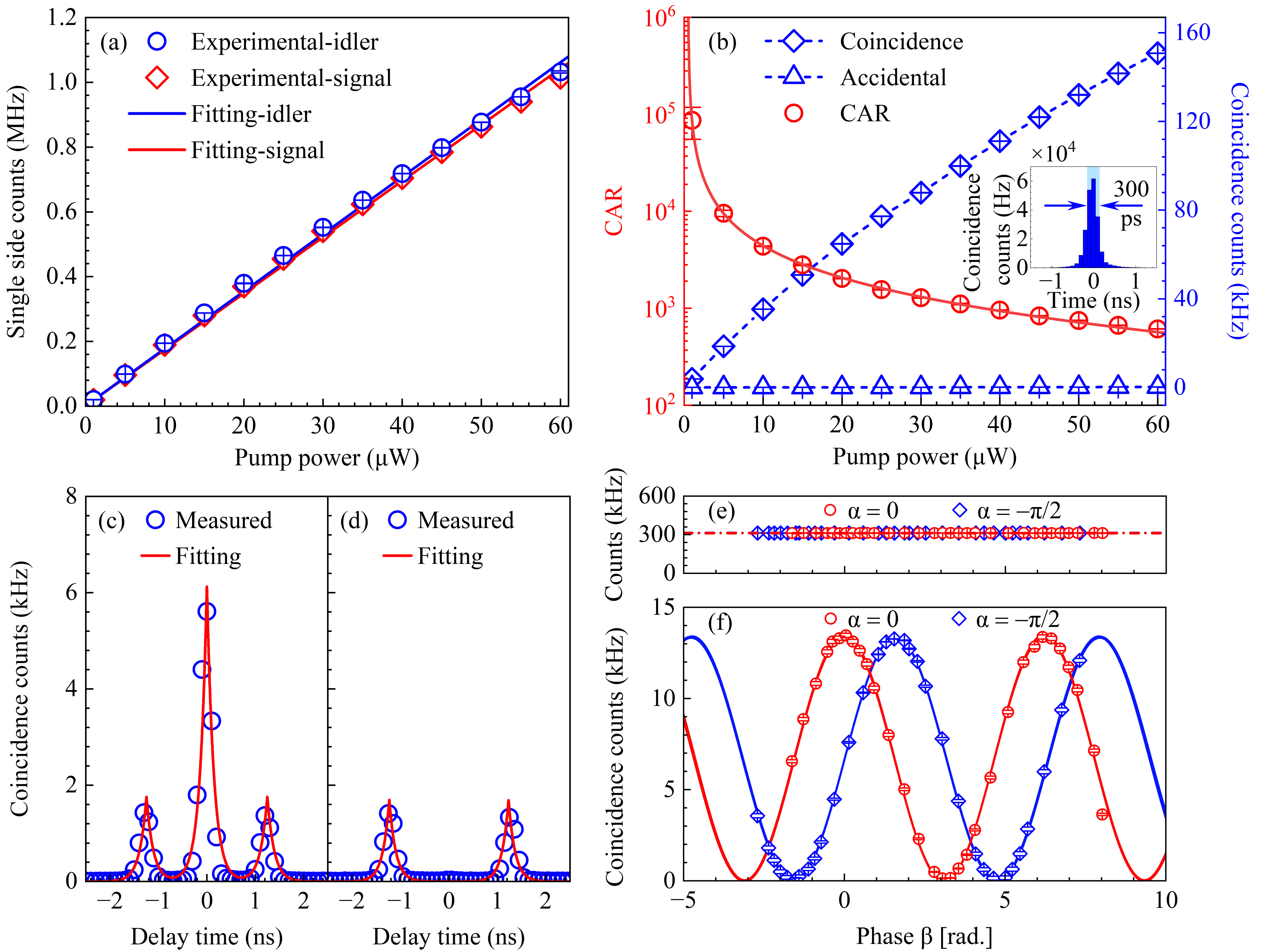}  
  \caption{
    Measurement results of the correlation properties and entanglement quality with the waveguide temperature at \SI{43.8}{\celsius}.  
    a) Single side count rates of the idler and signal photons versus pump power. 
    b) Coincidence count rate, accidental coincidence count rate, and CAR versus pump power. The inset shows the measured coincidence histogram of the idler and signal photons at a pump power of \SI{60} {\micro\watt}.
    c) Constructive Franson interference.
    d) Destructive Franson interference.
    e) Single side count rate of the idler photons in the Franson interferometer.
    f) Franson interference under two phase conditions: red circles for $\alpha = 0$, blue diamonds for $\alpha = -\pi/2$.}
  \label{fig:fig3}
\end{figure}

To characterize the performance of the quantum light source, we measure the correlation and energy-time entanglement properties using the experimental setups shown in Figure~\ref{fig:fig2}b and Figure~\ref{fig:fig2}c. The signal and idler photons generated in the PPLN waveguide are injected into a filtering system composed of DWDMs for spectral filtering, enabling precise frequency selection. Two DWDMs are employed to filter the idler and signal photons at wavelengths of \SI{1533.47} {\nano\meter} (C55 channel) and \SI{1530.33} {\nano\meter} (C59 channel), respectively. The DWDMs provide a sideband suppression ratio of \SI{120} {\decibel} and a filtering spacing of \SI{200} {\giga\hertz}, effectively suppressing non-target frequency components that could otherwise affect the correlation measurements. After filtering, the idler and signal photons are detected using superconducting nanowire single photon detectors (SNSPD), with detection efficiencies of \SI{85}{\percent} and \SI{80}{\percent}, respectively. The output signals of the SNSPDs are connected to a time-to-digital converter (TDC) for coincidence measurements. At a waveguide temperature of \SI{43.8}{\celsius}, \textbf{Figure~\ref{fig:fig3}}a shows the variation of the single side count rates of the idler and signal photons under different pump powers. At a pump power of \SI{60} {\micro\watt}, the count rates of the idler and signal photons reach \SI{1.03} {\mega\hertz} and \SI{1.01} {\mega\hertz}, respectively. Figure~\ref{fig:fig3}b presents the measured coincidence count rate, accidental coincidence count rate, and the CAR versus pump power. The inset of Figure~\ref{fig:fig3}b shows the measured coincidence histogram at a pump power of \SI{60} {\micro\watt}. The coincidence counts are collected within a \SI{300} {\pico\second} time window covering the coincidence peak, whereas the accidental coincidence counts are collected within another \SI{300} {\pico\second} time window far from the coincidence peak. At a pump power of \SI{60} {\micro\watt}, the coincidence count rate is \SI{150.81} {\kilo\hertz}. Using the corresponding single side count rates of the idler and signal photons, the corresponding PGR is calculated to be \SI{6.94} {\mega\hertz}, with a CAR of \num{610} $\pm$ \num{1}. When the pump power is reduced to \SI{1} {\micro\watt}, the CAR increases to \num{86661} $\pm$ \num{31559}, while the PGR remains at \SI{103.33} {\kilo\hertz}. These results demonstrate that the prepared quantum light source achieves a high PGR and high-quality correlation properties.

\FloatBarrier

The correlated photon pairs generated in the PPLN waveguide are sent to the setup shown in Figure~\ref{fig:fig2}c, where the energy-time entanglement is characterized using a Franson interferometer composed of a pair of identical unbalanced Mach-Zehnder interferometers (UMZI) under a pump power of \SI{60} {\micro\watt}. The signal and idler photons are first spectrally separated using a DWDM-based filtering system and then injected into two identical UMZIs with relative time delays of \SI{1.25} {\nano\second}. The coincidence histograms after interference are shown in Figure~\ref{fig:fig3}c and Figure~\ref{fig:fig3}d, corresponding to constructive and destructive Franson interference, respectively. Figure~\ref{fig:fig3}e shows the variation of the single side count rate of the idler photons during the Franson interference measurement. The red circles and blue diamonds correspond to the idler single side count rates under two different phase-scanning conditions ($\alpha = 0$ and $\alpha = -\pi/2$), respectively, while the red and blue dashed lines indicate the corresponding mean values. In both cases, the idler single side count rates remain stable at about \SI{312.70} {\kilo\hertz} and \SI{312.40} {\kilo\hertz}, indicating the absence of first-order interference during phase modulation. In Figure~\ref{fig:fig3}f, the red circles and blue diamonds represent the measured Franson interference data, while the red and blue curves show the fitting results obtained from a Monte Carlo method with \num{1000} repetitions. Without subtracting the accidental coincidence counts, the measured interference visibilities under the two phase settings $\alpha = 0$ and $\alpha = -\pi/2$ are (\num{99.93} $\pm$ \num{0.26})\% and (\num{99.96} $\pm$ \num{0.35})\%, respectively. The corresponding $S$-parameters are calculated as \num{2.826} $\pm$ \num{0.007} and \num{2.827} $\pm$ \num{0.010}, exceeding the classical limit by \num{118} and \num{83} standard deviations, respectively. These results significantly violate the Clauser-Horne-Shimony-Holt (CHSH) Bell inequality, indicating that the prepared quantum light source exhibits high-quality entanglement.

\section{Characterization of the Bandwidth Tunability of Broadband Entangled Photon Pairs}

\begin{figure}[htbp]
  \centering
  \includegraphics[width=1\linewidth]{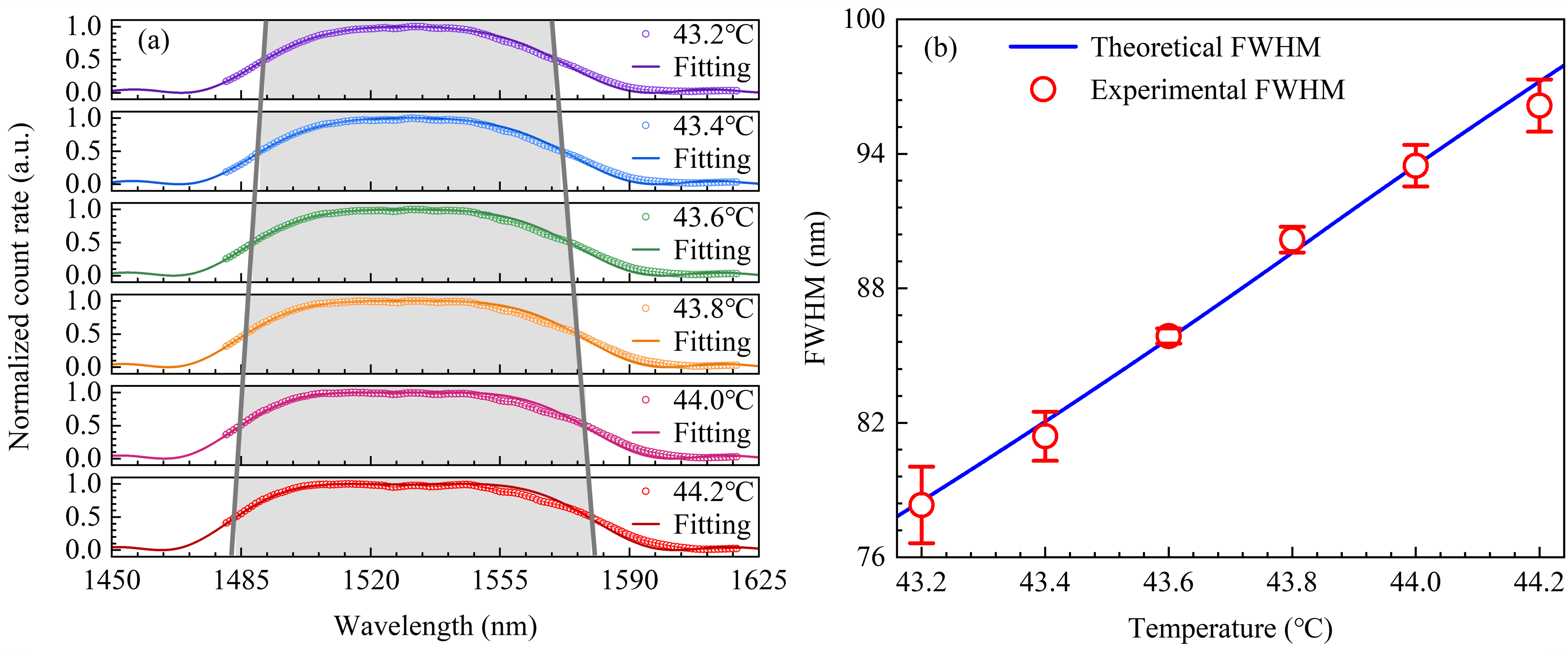}  
  \caption{
    Bandwidth tunability of correlated photon pairs. 
    a) Measured single photon spectra at different waveguide temperatures. 
    b) Theoretical and experimental spectral FWHMs at different waveguide temperatures.
  }
  \label{fig:fig4}
\end{figure}

To further characterize the tunability of the single photon spectral bandwidth of the quantum light source operating at 1.5 $\mu\text{m}$, we measure the single photon spectra of the correlated photon pairs by varying the waveguide temperature. As shown in Figure~\ref{fig:fig2}d, the correlated photon pairs generated in the PPLN waveguide are filtered by a tunable band-pass filter (TBF), selecting photons within the target wavelength range. The bandwidth of the TBF is set to \SI{650} {\pico\meter}, and its central wavelength is tuned from \SI{1480} {\nano\meter} to \SI{1620} {\nano\meter} with a step size of \SI{1} {\nano\meter}. These photons are then sent through a PC and detected using an SNSPD. The output signal of the SNSPD is connected to a TDC for single photon counting. The temperature of the TEC is set from \SI{43.2}{\celsius} to \SI{44.2}{\celsius}, with a step of \SI{0.2}{\celsius}. \textbf{Figure~\ref{fig:fig4}}a shows the normalized single photon spectra measured at different temperatures without subtracting the dark counts of the SNSPD. The circles in the figure represent the experimental data, while the curves correspond to the normalized fitting results obtained using Equation~\eqref{eq:rate}. The gray shaded region represents the full width at half maximum (FWHM) of the experimentally measured single photon spectrum, which is \SI[separate-uncertainty=true]{90.2 \pm 0.6} {\nano\meter} at a temperature of \SI{43.8}{\celsius}. The measurement results at the other temperatures are illustrated in Figure~\ref{fig:fig4}b. In the figure, the red circles indicate the experimentally measured FWHM, and the blue curve corresponds to the theoretical FWHM. As the temperature increases, the FWHM of the single photon spectrum increases from \SI{78.3} {\nano\meter} to \SI{96.2} {\nano\meter}, showing an increasing trend with temperature and consistency with the theoretical curve.

\FloatBarrier

Furthermore, we characterize the correlation and energy-time entanglement properties of the quantum light source operating at different bandwidths. The measurement results are summarized in \textbf{Table~\ref{tab:tabular}}. Over the bandwidth-tuning range corresponding to a temperature variation of \SI{1}{\celsius}, the single side count rates of both the signal and idler photons remain above \SI{0.94} {\mega\hertz}, and the corresponding PGRs are no less than \SI{6.33} {\mega\hertz}. The CARs remain above \num{608} under all temperature conditions, indicating that the source exhibits good correlation properties across different bandwidths. The corresponding Franson interference visibilities exceed \num{99.06}\% under both phase settings $\alpha = 0$ and $\alpha = -\pi/2$, and show no noticeable degradation with temperature. Combined with the CHSH Bell inequality violation results listed in Table~\ref{tab:tabular}, the generated entangled photon pairs maintain high-quality entanglement properties under different bandwidth conditions. These results demonstrate that, by controlling the operating temperature of the PPLN waveguide, predictable tuning of the single photon spectral bandwidth can be achieved while preserving high-quality entanglement properties.

\begin{table}[htbp]
  \caption{Experimental results of the correlation properties and entanglement quality of the bandwidth-tunable quantum light source at different temperatures.}
  \label{tab:tabular}
  \centering

  {%
  
  \rowcolors{1}{white}{white}

  \arrayrulecolor[RGB]{50,50,50}
  \setlength{\arrayrulewidth}{1.5pt}

  \resizebox{\textwidth}{!}{%
    \begin{tabular}{ccccccccc}
      \hline
      \makecell[c]{Temperature \\ {[\si{\celsius}]}}
      & \makecell[c]{C55 \\ {[\si{\mega\hertz}]}}
      & \makecell[c]{C59 \\ {[\si{\mega\hertz}]}}
      & \makecell[c]{PGR \\ {[\si{\mega\hertz}]}}
      & CAR
      & \makecell[c]{Visibility {[\si{\percent}]} \\ ($\alpha = 0$)}
      & \makecell[c]{Violation of \\ Bell inequality}
      & \makecell[c]{Visibility {[\si{\percent}]} \\ ($\alpha = -\pi/2$)}
      & \makecell[c]{Violation of \\ Bell inequality}
      \\ \hline

      43.2 & 0.96 & 0.94 & 6.33 & 675 $\pm$ 4 & 99.11 $\pm$ 0.19 & 161 & 99.38 $\pm$ 0.19 & 162 \\
      43.4 & 1.02 & 1.00 & 6.78 & 625 $\pm$ 1 & 99.22 $\pm$ 0.25 & 115 & 99.06 $\pm$ 0.16 & 160 \\
      43.6 & 1.04 & 1.02 & 6.93 & 609 $\pm$ 2 & 99.84 $\pm$ 0.22 & 137 & 99.24 $\pm$ 0.16 & 161 \\
      43.8 & 1.03 & 1.01 & 6.94 & 610 $\pm$ 1 & 99.93 $\pm$ 0.26 & 118 & 99.96 $\pm$ 0.35 & 83 \\
      44.0 & 1.04 & 1.02 & 6.97 & 608 $\pm$ 2 & 99.72 $\pm$ 0.21 & 137 & 99.83 $\pm$ 0.22 & 137 \\
      44.2 & 1.03 & 1.01 & 6.90 & 616 $\pm$ 5 & 99.80 $\pm$ 0.20 & 137 & 99.79 $\pm$ 0.16 & 164 \\
      \hline
    \end{tabular}%
  }%
  }
\end{table}

\section{Conclusion}

In this work, we have demonstrated a bandwidth-tunable quantum light source operating at 1.5 $\mu\text{m}$ based on a single PPLN waveguide, and have investigated the tuning characteristics of its single photon spectral bandwidth as well as the corresponding correlation and entanglement properties. The spectral bandwidth of the broadband correlated photon pairs has been tuned by controlling the temperature of the waveguide, while maintaining megahertz-level PGRs and the energy-time entanglement property. Our results have shown that the bandwidth of the quantum light source can be tuned from \SI{78.3} {\nano\meter} to \SI{96.2} {\nano\meter} within \SI{1}{\celsius} temperature in agreement with theoretical calculations. The correlated PGRs remain no less than \SI{6.33} {\mega\hertz} with CARs exceeding \num{608}. The Franson interference visibilities are higher than \num{99.06}\%. Our method provides a feasible approach for developing high-performance broadband quantum light sources, laying the foundation for large-scale quantum networks.

\section*{Acknowledgements} \par 

This work was supported by Quantum Science and Technology—National Science and Technology Major Project (Nos. 2024ZD0300800 and 2021ZD0301702), Sichuan Science and Technology Program (Nos. 2024YFHZ0370, 2024YFHZ0368,  2024YFHZ0369, 2026NSFSC1439), National Natural Science Foundation of China (Nos. 62475039, 62405046, 62375043), and Tianfu Jiangxi Laboratory (No. TFJX-ZD-2025-005).

\section*{Conflict of Interests} \par 
The authors declare no conﬂict of interest.

\section*{Data Availability Statement} \par 
The data that support the findings of this study are available from the corresponding author upon reasonable request.

\medskip


\medskip

\end{document}